# AUTHOR RESPONSES TO IEEE ACCESS SUBMISSION QUESTIONS

| | |
|---|---|
| Author chosen manuscript type: | Research Article |
| Author explanation /justification for choosing this manuscript type: | Our previously  manuscript Access-2022-13685 has not been recommended for publication in IEEE Access in its current form. However, editor(s) encouraged us to address two reviewers' concerns through resubmission. This manuscript is a resubmission's version updated according to the reviewer's comments. |
| Author description of how this manuscript fits within the scope of IEEE Access: | Some of my papers have been published in IEEE ACCESS: N. Ahn and D. H. Lee, "Schemes for Privacy Data Destruction in a NAND Flash Memory," in IEEE Access, vol. 7, pp. 181305-181313, 2019, doi: 10.1109/ACCESS.2019.2958628. N. Y. Ahn and D. H. Lee, "Forensics and Anti-Forensics of a NAND Flash Memory: From a Copy-Back Program Perspective," in IEEE Access, vol. 9, pp. 14130-14137, 2021, doi: 10.1109/ACCESS.2021.3052353. N. Y. Ahn and D. H. Lee, "Forensic Issues and Techniques to Improve Security in SSD With Flex Capacity Feature," in IEEE Access, vol. 9, pp. 167067-167075, 2021, doi: 10.1109/ACCESS.2021.3136483. |
| Author description detailing the unique contribution of the manuscript related to existing literature: | Secure deletion is becoming a very important issue in NAND flash memory. My paper introduces the necessity of a verification technique applicable to numerous security deletion techniques, and a verification method and device for the first time. It is expected that these techniques will be discussed a lot in the future. |
| | |

# Security of IoT Device: Perspective Forensic/Anti-Forensic Issues on Invalid Area of NAND Flash Memory

Na Young Ahn[1], and Dong Hoon Lee[2], Member, IEEE

[1] Institute of Cyber Security & Privacy, Korea University, Seoul, South Korea
[2] Graduate School of Information Security, Korea University, Seoul, South Korea

Corresponding author: Dong Hoon Lee (e-mail: donghlee@ korea.ac.kr).

**ABSTRACT** NAND flash memory-based IoT device can potentially still leave behind original personal data in an invalid area even if the data has been deleted. In this paper, we raise the forensic issue of original data remaining in unmanaged blocks caused by NAND flash memory and introduce methods for secure deletion of such data in the invalid area. We also propose a verification technique for secure deletion that is performed based on cell count information, which refers to the difference in bits between personal data and data stored in the block. The pass/fail of the verification technique according to the cell count information is determined in consideration of error correction capabilities. With the forensic issue of de-identification being a vital theme in the big data industry, the threat of serious privacy breaches coupled with our proposal to prevent these attacks will prove to be critical technological necessities in the future.

## I. INTRODUCTION

In general, de-identification of personal information is a critical issue in big data ecosystems [1-3]. In order to enhance privacy, de-identification technologies are required to reduce the threat of re-identification [4], about which there is active discussion specifically surrounding single output, linkability, inference, and indistinguishability [5]. Single output refers to the degree to which a set corresponding to a specific individual can be identified in the entire data set [6]. Linkability refers to the degree to which individualized information is identified as information of a specific individual through linking with other information [7]. The possibility of inference refers to the degree to which a specific individual is inferred through the attribute value of specific information [8]. Indistinguishability refers to the degree to which a specific information value is distinguished from a specific individual included in a specific group or affiliation [9]. In general, de-identification techniques are introduced in various ways, such as masking, pseudonymization, anonymity, diversity, similarity, sampling, and aggregation [10, 11]. In order to design the privacy of big data, the European Union presents requirements suitable for the four stages of data collection, data analysis, data storage, and data use [12].

Data-hiding NAND flash memory is a nonvolatile memory having features of low power consumption and high performance read/write operations [13]. For this reason, the use of NAND flash memory in storage devices, such as smart phones, IoT devices, and SSDs, is explosively increasing. In NAND flash memory, a unit of a write operation (or a program operation) and an erase operation are fundamentally different. In addition, it is generally known that NAND flash memory cannot be overwritten [14]. For this reason, NAND flash memory performs various internal operations to extend life span and improve performance. Related to these internal actions, studies on the dangers of forensics have recently been raised [15-21].

This study discusses a method of verifying forensic possibility, complete deletion, and complete deletion when de-identification processing operations are performed in NAND flash memory-based IoT devices. Chapter 2 describes how to define the original data in NAND flash memory with an error correction function, and Chapter 3 discusses the forensic targets of NAND flash memory. In Chapter 4, we address the forensic issue of de-identification of NAND flash-based storage devices. Even after processing the de-identification processing operation, the personal data existing in the unmanaged block will be described in detail. Chapter 5 thus proposes a verification method for such secure deletion after completely deleting personal data existing in the unmanaged block caused by the de-identification operation. The proposed verification method is a method of One-to-One matching and comparing data stored in a block with personal data. Finally, Chapter 6 shows the results of the success and failure of secure deletion according to the above verification technique. This study, for the first time, raises the issue of remaining original data caused by the de-identification processing operation of NAND flash memory-based storage devices. In addition, it confirms that the complete de-identification processing operation necessarily requires verification of the secure deletion of such residual data.

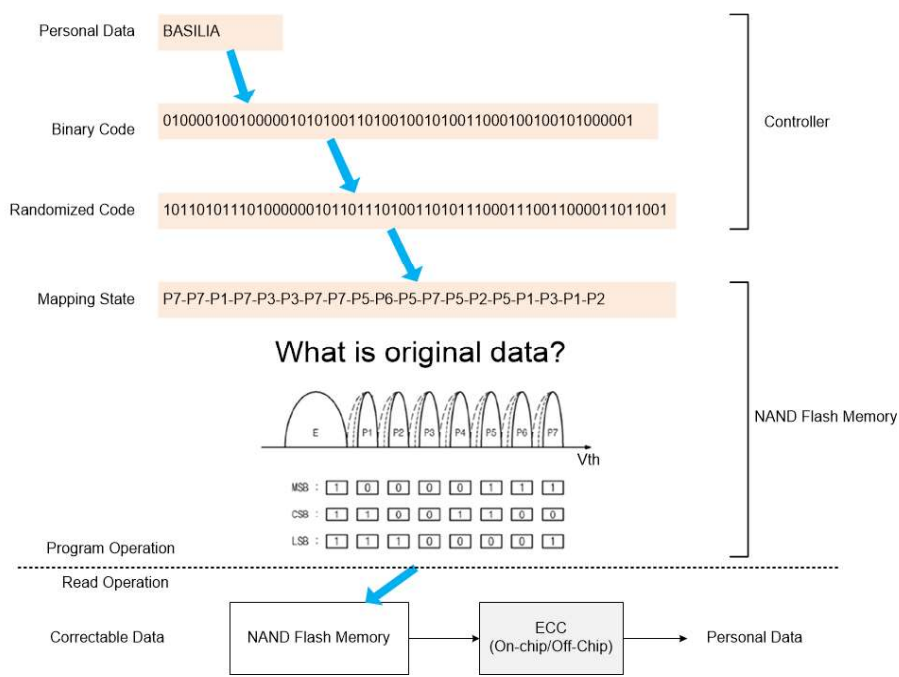

FIGURE 1. Original data from the lifetime of personal data in NAND flash memory, original data during a program operation, and original data during a read operation. The original data includes correctable data.

While the IoT environment introduces great convenience, at the same time, it can also lead to serious privacy breaches [22, 23, 44]. Our study raises a forensic issue for IoT devices based on NAND flash memory. In addition, by introducing realistic verification techniques for secure deletion, our research will be of great help to chip manufacturers and users in finding answers to upcoming or already occurring problems.

## II. ORIGINAL DATA IN NAND FLASH MEMORY

Digital forensics is a generic term for forensic techniques used for investigation by analyzing digital evidence. Recently, in connection with IoT devices, there is abundant research on digital forensics for NAND flash memory [15-21], which poses new questions about what the target of digital forensics in NAND flash memory is. Before defining this, though, it is first necessary to digitally grasp the concept of what original data is. The process of storing data in NAND flash memory is as follows. As shown in Fig. 1, the personal data stored in NAND flash memory in the controller is converted into binary code. Thereafter, the converted binary code is randomized (or scrambled) based on a random (scramble) algorithm. Thereafter, the randomized data is state mapped to be stored in NAND flash memory [17, 24]. A threshold voltage Vth corresponding to the state-mapped state is programmed in NAND flash memory. For example, the binary code corresponding to personal data BASILIA is 01000010010000010101001101001001010011000100100101000001. Is the binary code original data? This binary code is randomized (or scrambled) to 57 bits as 101101011101000000101101110100110101110001110011000011011001. Then, is this random data original data?

The mapping states corresponding to the 57-bit random data described above are P7-P7-P1-P7-P3-P3-P7-P7-P5-P6-P5-P7-P5-P2-P5-P1-P3-P1-P2. That is, through the program operation, each of the 19 memory cells has a threshold voltage Vth in a corresponding state. Then, is the state programmed in each of the 19 flash memory cells original data? It is necessary to define all personal data, binary code, random data, and mapping state of memory cells as original data for digital forensics. In the real world, personal data BASILIA is the mapping state of memory cells P7-P7-P1-P7-P3-P3-P7-P7-P5-P6-P5-P7-P5-P2-P5-P1-P3-P1. Regarding ASCII code conversion algorithm, random algorithm, and mapping table, data BASILIA is the mapping state of memory cells, which is the same as P7-P7-P1-P7-P3-P3-P7-P7-P5-P6-P5-P7-P5-P2-P5-P1-P3-P1-P2. Moreover, since NAND flash memory basically includes an error correction circuit internally or performs an error correction operation in an external controller, the number of mapping states of equivalent memory cells can be greatly increased by the number of bits capable of error correction. This problem necessitates a separate definition of original data in NAND flash memory. That is, it is pertinent to recognize even data capable of error correction as original data.

In general, in a read operation of NAND flash memory, even if as many errors exist as the number of bits determined in advance based on an error correction circuit, original data can be restored. For this reason, even if the mapping states of the memory cells do not match exactly, if personal data can be restored, we need to recognize these mapping states as equivalent original data.

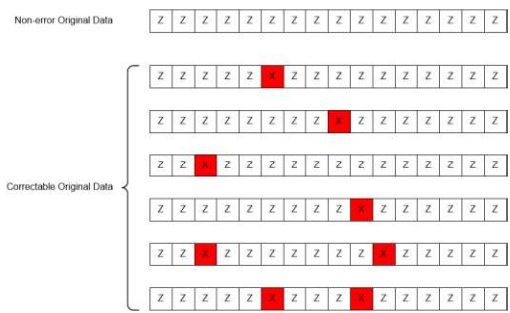

**FIGURE 2.** Correctable original data. There is non-error original data and much original data that has been corrected although errors exist.

Fig. 2 exemplarily shows original data without errors and originals capable of correcting errors. If there is an error of 1 bit or 2 bits and error correction is possible, these data pieces are all original data that can be error corrected. Previously, the concept of original data was defined as exactly matching bits, but from now on, original data stored in NAND flash memory needs to be defined differently. In NAND flash memory, original data can be defined as data that stores other data that can be error corrected by an on/off-chip error correction circuit [25, 26]. However, the problem is that there may be a number of such error-correctable data in NAND flash memory.

## III. FORENSIC TARGET OF NAND FLASH MEMORY

In general, NAND flash memory includes a plurality of memory blocks, referring to Fig. 3. Among this plurality of memory blocks, there are both managed memory blocks and unmanaged memory blocks. Each memory block includes pages corresponding to word lines, and each page includes a plurality of memory cells connected to a corresponding word line. Each memory cell is implemented to store at least one bit. In general, program/read operations are performed in one-page units, an erase operation is performed in one-block units. The units of program/read operation and erase operation are different [13]. Also, it is known that NAND flash memory cannot be overwritten by default [14]. These features increase the complexity of block management [27, 28].

### A. Block Management for Reliability

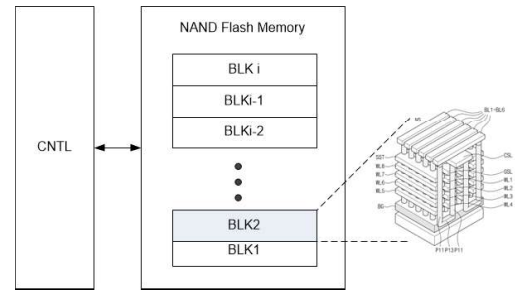

**FIGURE 3.** Memory blocks of NAND flash memory.

In NAND flash memory, various block management operations are internally performed to improve the reliability and lifetime of blocks. Representatively, such block management operations include a wear leveling operation and a garbage collection operation, referring to Fig. 4 [29 - 31].

Wear leveling refers to the operation of managing the wear level of a block to a certain level. Each of the memory blocks is wear leveled by PE cycles [29]. Here, PE cycles means the number of times a program operation and an erase operation have been performed. Wear leveling refers to an operation of replacing a corresponding block with a free block when the PE cycles of one block exceeds a certain level. In wear leveling, an invalidation block may be replaced with a free block in a free space.

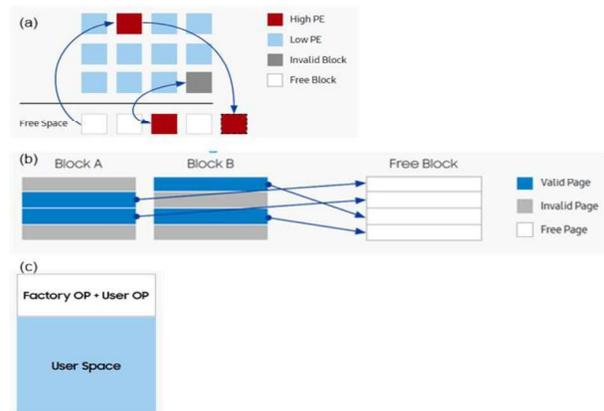

**FIGURE 4.** (a) Wear leveling (b) Garbage collection (c) Over provisioning.

The block includes a plurality of pages. Each page may be either valid or invalid data. Garbage collection involves collecting valid pages of different blocks and creating a new block [30]. At this time, the original blocks A and B are invalidated. In the garbage collection operation, blocks A and B storing original data are generally treated as an over-provisioning area OP [32]. The user cannot access the OP original blocks. The flash translation layer of the controller manages access to the managed memory blocks based on addresses received from the host. In addition, the controller performs bad block management on unmanaged memory blocks, but it cannot access them based on addresses received from the host. Therefore, users of NAND flash memory cannot access unmanaged blocks. However, a hacker who has secured map table information and bad block information can access unmanaged blocks by a simple method.

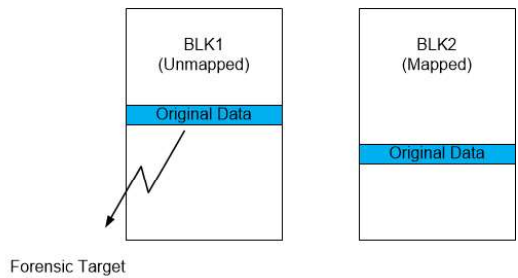

**FIGURE 5.** Forensic target in invalid area of NAND flash memory. The target of hackers is the original data of unmapped BLK1.

### B. Limitation of TRIM Operation

A TRIM operation is performed inside the storage device when the user deletes data [16, 33], and it can accurately erase pages in a logically invalidated file. However, since the TRIM operation basically uses a block-erase operation, it weakens the deterioration characteristic of the memory block and is not desirable in terms of management, such as power consumption due to an additional erase operation. In general, it is known that the erase operation consumes a relatively large amount of power compared to the program/read operation. NAND chip manufacturers who want to guarantee 1,000 erase cycles as a guarantee will simply design this TRIM operation to be infrequent. As a result, even though the original data of the storage device is deleted or updated by the user, it is very likely that the previous original data still exists in at least one block of NAND flash memory.

### C. Scrubbing/Partial Programming

The necessity of complete erasure in NAND flash memory is a topic of continuous discussion among researchers, and a NAND flash memory-based complete erasure technique was first introduced in 2017 [15]. In this groundbreaking research, Ahn proposed the need to apply a duty to delete from NAND flash memory. It is known that the existing NAND flash memory cannot be overwritten, but it is also known that there are states that can be partially overwritten in a multi-bit program operation. Ahn proposed an erasure technique that partially programmed random data into these states. The deletion technique using this partial program is advantageous in terms of cost compared to the existing TRIM technique, which is based on block erasure. In 2018, Wang et al. proposed a clean erasure technique using scrubbing technology [34]. Here, the scrubbing technique forces the threshold voltage of a deteriorated memory cell to be higher than a certain level. For example, all threshold voltages of memory cells are programmed above a certain level by a kind of program operation that affects all data, for example, '1' bit. However, in 2020, Hasan et al. revealed that this scrubbing technique is partially recoverable [35]. Hasan proposed performing a partial program operation in the scrubbing technique for secure deletion. In 2019, Ahn et al. proposed three techniques for secure deletion: a partial program technique, a down-bit program technique, and a deletion pulse technique [17]. In 2020, Ahn mentioned the recovery operation to minimize the victim of adjacent pages according to the secure deletion technique [20]. Lin et al. introduced fast sanitization using zero live data [36]. Meanwhile, studies to achieve secure deletion using cryptographic techniques are also ongoing [37, 38].

### D. Forensic Target in NAND Flash Memory

The target of digital forensics may be the original data stored in the unmanaged block of invalid area, referring to Fig. 5. Here, an unmanaged block is a block that stores valid data but cannot be accessed by the user, a block that stores invalid data but is accessible by the user, or a block that is not accessible by the user regardless of whether the data is valid. In addition, the invalid data here is referred to as data that can be restored to original data using a simple technique. These unmanaged blocks may exist in the user space or in the OP space. At this juncture, the controller knows the physical address of the unmanaged block. This means that the controller can replace it with a managed block at any time. Thus, the original data stored in the currently unmanaged block can be retrieved from the controller at any time. Only the original data of the normally managed block is considered a forensic target, which is half the answer. NAND flash memory contains numerous unmanaged blocks for efficient management and performance improvement. As integration increases and the number of memory cells with low reliability increases, the number of such unmanaged blocks will increase steadily. This is the point at which a detailed study on the forensic target of original data stored in unmanaged blocks is necessary.

We have established that there is not just one original data stored in NAND flash memory. Indeed, original data stored in unmanaged blocks inevitably exists. Ahn and Lee, in reference to original data remaining in the unmanaged block according to the copy-back program operation, suggested that original data remains in unmanaged blocks due to garbage collection. In addition to these operations, NAND flash memory performs various internal operations or background operations to improve data reliability. These actions will positively improve the reliability of the data, but the implication is that the number of unmanaged original data is growing. In particular, to improve the performance of recent storage devices, a relatively large portion of OP space must be used.

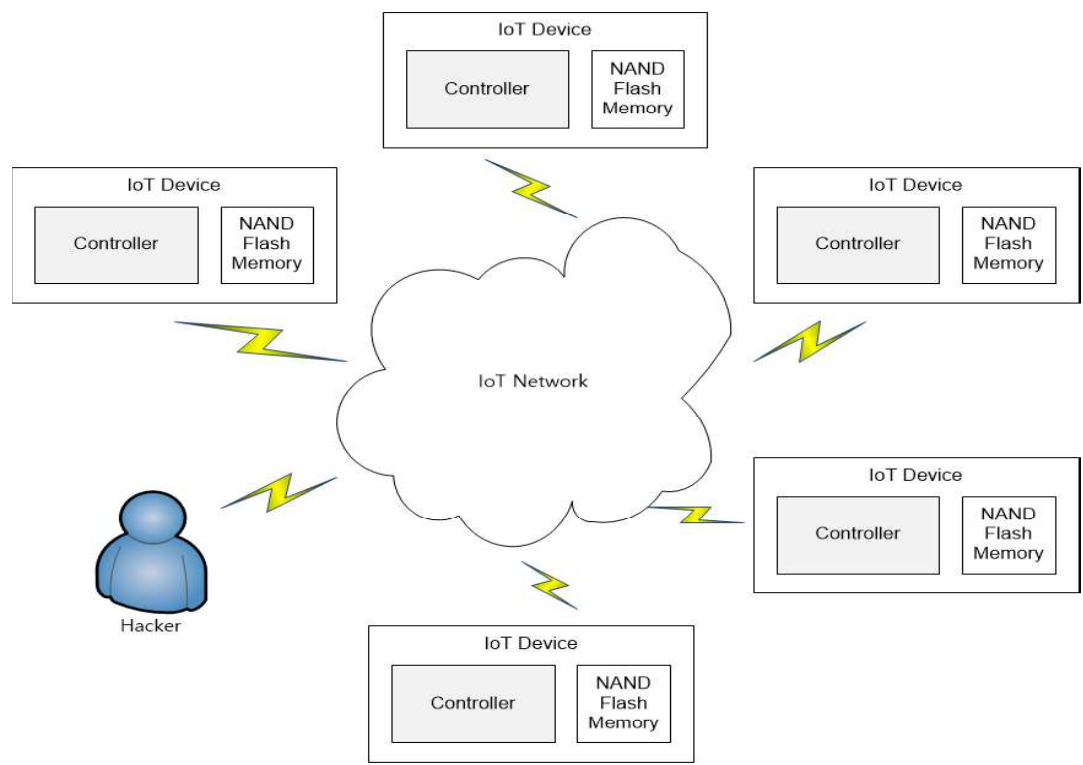

**FIGURE 6.** Hacking threat of IoT devices based on NAND flash memory.

Previous forensic studies of NAND flash memory focused on memory stored in managed blocks. Hasan et al. mentioned the possibility of restoring data even if data is deleted via a scrubbing operation. However, as Ahn and Lee's previous studies show, the original data issue for unmanaged blocks can be fatal. Unmanaged blocks are the only areas not visible to users of the storage device. This is because unauthorized access to unmanaged blocks is quite easy if the map table is freely adjustable in the controller, and forensics targets the original data included in these unmapped blocks. Forensics for unmanaged blocks is possible without separating NAND flash memory from the controller, which can be even more fatal because it implies that forensics can be easily performed on NAND flash memory by hacking the controller. Additionally, hackers can read original data from IoT devices through IoT networks without physical access.

Hackers that gain control of an IoT device controller can, at any time, steal personal information from unmanaged blocks of NAND flash memory, referring to Fig. 6. Users may think that they have completely erased their personal information from their IoT devices, but hackers can acquire personal information despite an understandable naïveté on the part of the user. Although we are only addressing this fact now in this paper, hackers may already be performing such malicious acts, which are critical and gravely problematic. Storage devices based on NAND flash memory are essential components in IoT devices. The IoT environment is a vast ecosystem, and malicious actors have the potential to illegally acquire personal information from this ecosystem and commit more serious crimes using the acquired data.

## IV. FORENSIC ISSUES OF DE-IDENIFICATION IN SSDs

De-identification is a process used to prevent personal information from being disclosed, and this process is important from a data privacy protection point of view. De-identification operations are used in communications, multimedia, biometrics, big data, cloud computing, data mining, and social network audio-video surveillance [39, 40]. Recently, de-identification information has been classified as personal information as long as there is legal precedent [12]. It should be noted that there is active discussion and research on de-identification techniques that take into account this binding possibility [4-10]. In general, de-identification techniques are divided into encryption techniques, deletion techniques, pseudonymization techniques, dissection techniques, and randomization techniques.

The point we are presently interested in is the physical storage device that performs de-identification operations. In a database using NAND flash-based storage devices, the current de-identification operation is a very dangerous processing operation. This is because there are too many issues of remaining original data in NAND flash memories. The de-identification operation is usually performed the following way: personal information stored in the storage device is read. Then, the read personal information is de-identified using a de-identification technique. Finally, the de-identified information is stored back on the storage device. Unfortunately, even though the de-identified data is updated to the storage device, the existing personal information is not automatically deleted from the device. Currently, there is a high possibility that there are both blocks in which personal information is stored and blocks in which de-identification data is stored. This is because the de-identified data is not overwritten with personal information from the existing block; rather, it is programmed in a new block. Although an attempt was made to minimize the exposure of personal information through the de-identification operation, the original form of personal information remains somewhere on the storage device.

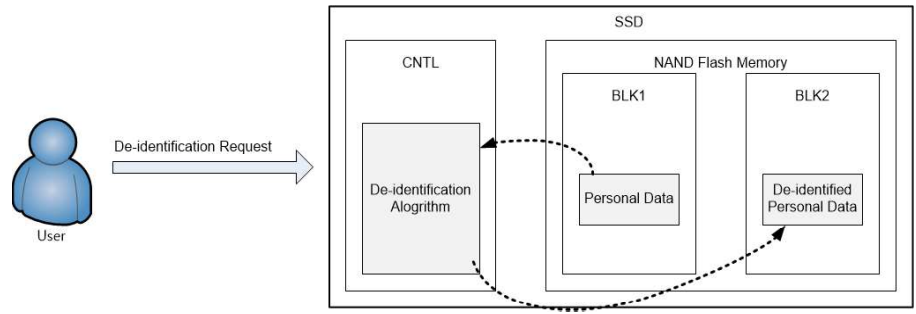

**FIGURE 7.** De-identification Process in SSD based on NAND flash memory.

*A. Residual data issues in de-identification operation*

As shown in Fig. 7, personal information is still stored in the first block, and de-identification data is stored in the second block. There is a block in which personal information is stored in the storage device, but the user cannot access it by the mapping table. However, the mapping table can be activated at any time. Therefore, although personal information is said to be de-identified, in reality it is not de-identified at all within the storage device. Accordingly, in a storage device based on NAND flash memory, an additional operation is required for personal information that cannot be identified in the de-identification operation.

*B. Secure Deletion of Personal Data using Scrubbing*

Personal information is de-identified and the de-identified data is stored on a storage device, but the de-identification operation does not end there. More steps are needed to completely delete personal information stored in the existing block, namely an erase operation for the existing block or a partial erasure operation for a page having the personal information of the existing block. Detailed operations for partial deletion have already been introduced by Ahn and Lee [17, 20, 21]. Generally speaking, NAND flash memory does not support the update operation. This means that, even if existing de-identification technologies are applied, there is an issue of original personal information handling. Therefore, we propose a de-identification technology without such original personal information handling issues. Our de-identification technique is similar to the update operation except that the concept is only slightly different. It performs a scrubbing operation on the page containing the original personal information, referring to Fig. 8. Through this scrubbing operation, the data stored in the page is entirely changed to random data, and restoration is impossible.

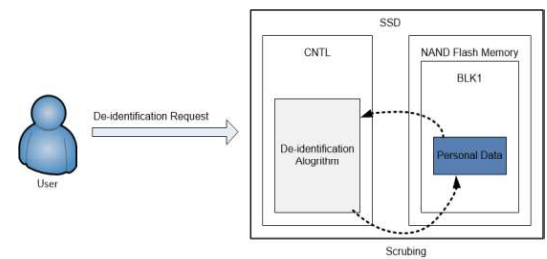

**FIGURE 8.** Secure deletion using scrubbing technology. SSD is connected to or embedded in IoT devices

*C. Necessity of Verification for Secure Deletion*

Technologies for secure deletion in NAND flash memory are popular research topics in recent years [34-38], but the issue of secure deletion is one that should be further considered by the chip manufacturer. However, the most important item that needs to be verified is whether or not the secure deletion operation completes successfully. It seems like a way to look silly, but the most reliable verification method is to read the data stored in all physical blocks and determine whether there is identifiable personal information among the read data. However, such approach can never be recommended in reality because the time and the cost of the verification operation are enormous. There is a significant need for research on a method that can perform verification at a low cost and in a short amount of time using the unique characteristics of NAND flash memory-based storage devices. In other words, an efficient verification scheme for secure deletion is greatly needed. Such verification must be freely possible in the controller, and real-time verification must be possible for all physical blocks.

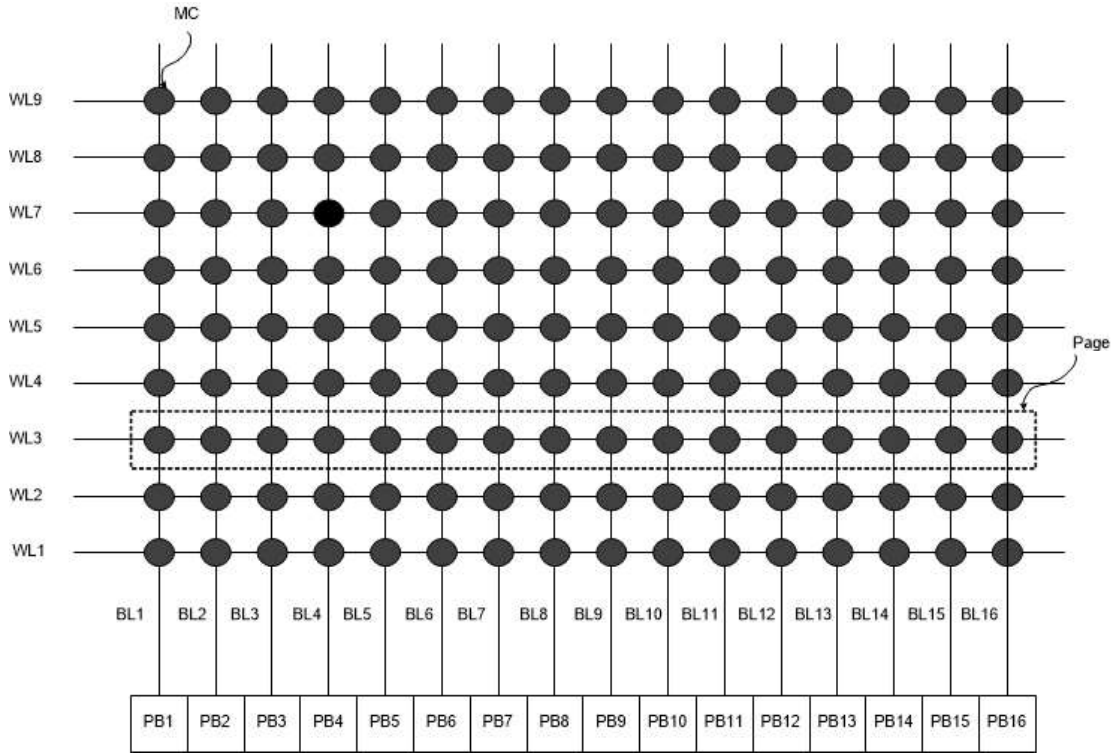

**FIGURE 9.** General Memory core of NAND flash memory.

## V. VERIFICATION SCHEME OF SECURE DELETION

Before introducing our method of verifying secure deletion, we discuss the core of NAND flash memory, which is helpful to understand our proposed secure deletion verification. This core is composed of a plurality of pages corresponding to a plurality of word lines and a plurality of page buffers connected to each of the pages by bit lines. Memory cells MCs are disposed at intersections of word lines WL1 to WL9 and bit lines BL1 to BL16, referring to Fig. 9. In each of the plurality of bit lines BL1 to BL16, the plurality of page buffers PB1 to PB16 sense data in a read operation or store data to be programmed in a program operation. Each of the page buffers PB1 to PB16 temporarily stores data for programming data in a memory cell in a program operation or temporarily stores data sensed from a memory cell in a read operation. Each of the page buffers PB1 to PB16 determines a bit of data according to a voltage of a corresponding bit line. Each of the page buffers PB1 to PB16 includes a latch storing sensing data, a latch storing data to be programmed, and at least one latch temporarily storing data that is necessary for operation.

In general, program/read operations are performed in a page unit composed of memory cells connected to a selected word line. One page may consist of a plurality of sectors. NAND flash memory can also perform program/read operations in units of sectors. Our proposed verification operation for secure deletion may be performed in a unit of a page or in a unit of a sector. Depending on the size of the personal information to be compared, the verification unit may vary.

*A. One-to-One Matching Comparison in Page Buffer*

Each page buffer includes a plurality of latches. As described above, the page buffer includes a latch that stores data sensed from a memory cell and a latch that stores data received from the outside. The page buffer can compare data read from a memory cell of a block to be verified and data corresponding to personal information received from the outside on 1:1 matching. The page buffer may compare sensing data and received data and output a bit corresponding to the comparison result.

*B. Determining Equal vs. Unequal Items*

If the values of the comparison bits output from the plurality of page buffers are all the same, data corresponding to personal information is still stored in the block. As such, it cannot be determined that personal information no longer exists even if all the bit values output from the page buffers are the same. As mentioned earlier, if some bits are different but error correctable, they are not identical but should be considered equal. The verification operation of secure deletion should fundamentally consider error correction capabilities. This is because, if correctable by the ECC circuit, it must be determined to be equivalent. Therefore, the criteria for determining verification are not the same, but whether or not they are equal.

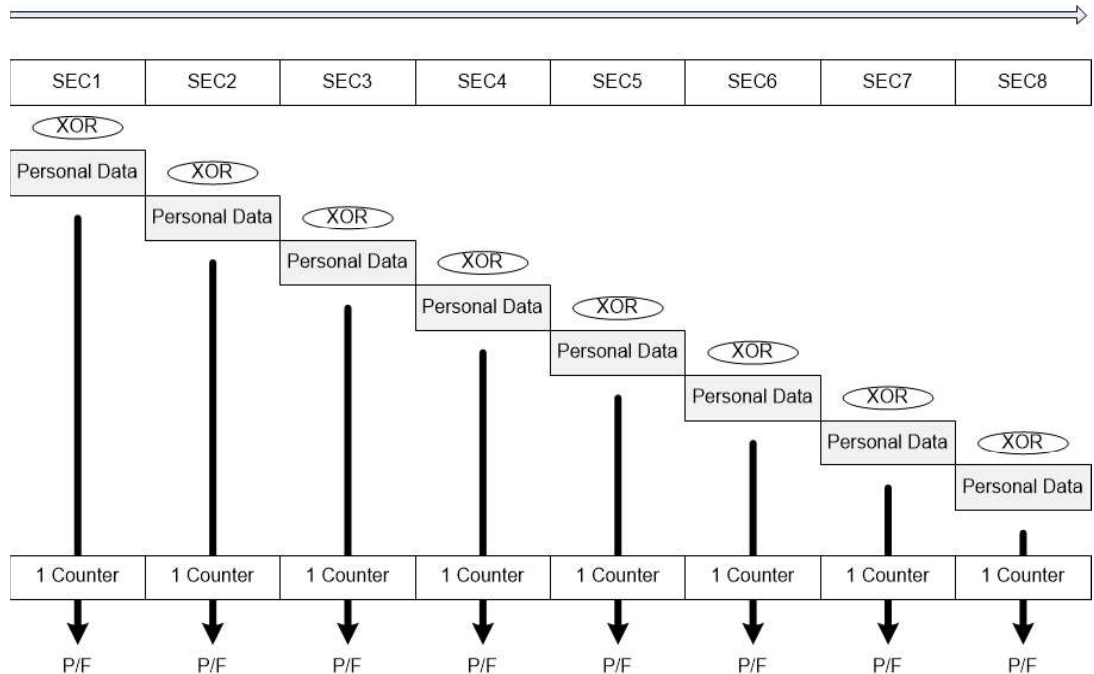

**FIGURE 10.** Verification logic for secure deletion using cell counters. XOR operations are performed on the data sensed by each sector and personal data. 1-bit counters count the results of the operations.

### C. Cell Counter

NAND flash memory generally has a cell counter that counts the number of memory cells forming a current path in a channel in response to a read voltage [41]. Such cell count information is used to determine if a program operation is successful or to check the deterioration state of a cell. This cell counter can be used to verify the proposed secure deletion. By simply changing the control signals of the page buffer, the cell counter can count the XOR-calculated value of the sensed data and the data corresponding to the received personal data, referring to Fig. 10. The count information of the XOR-operated values can be used to determine whether the data stored in the block and personal information are equal or not. That is, this count information indicates how many bits are different when comparing the personal data with the data stored in the block. If the count information is less than or equal to the number of errors correctable, the data stored in the block is personal data. In this case, it is determined that the de-identification operation has failed. On the other hand, if the count information is greater than the number of levels capable of error correction, the data stored in the block is not personal data. In this case, it is deemed that the de-identification operation has been performed successfully.

### D. Page Buffer with XOR Logic

The abovementioned verification technique requires XOR logic to be added to the existing page buffer. In basic terms, XOR logic outputs different input values. For example, if a data bit sensed from a memory cell and a bit corresponding to personal data are different from each other, the XOR logic outputs data "1." On the other hand, if the data bit sensed from the memory cell and the bit corresponding to the personal data are the same, the XOR logic outputs data "0." Adding XOR logic to each of the page buffers can be an area burden. However, this XOR logic can replace the complex latch operation of a program operation, an erase operation, and a read operation with a simple XOR logic. This discussion is left to the next generation of researchers. The proposed verification technique can quickly check the success or failure of secure deletion within the block using only the count information of the cell counter.

### E. Failure/Success of De-identification Operation

After the de-identification operation, it is necessary to verify whether or not secure deletion is possible. NAND flash memory has the potential to contain one or more pieces of original data. Therefore, NAND flash memory needs to perform a verification operation after applying the secure deletion technique. The success of the de-identification operation is determined by the success of secure deletion of an unmanaged block. The proposed verification technique uses an operation of XORing sector data and personal data, referring to Fig. 10. XOR operation is performed on the data read from the sector of the page and personal data, and the number of bits "1" is counted according to the execution result. If the counted value is greater than a predetermined value, the secure deletion operation is determined as having passed. Here, the predetermined value may vary according to the error correction capability of the error correction circuit. As shown in Fig. 10, XOR operations are sequentially performed on eight sectors of a page, and a pass or fail of the complete erasure technique may be determined according to a counting value of bit "1," depending on the result.

In summary, the de-identification process involves reading personal data from a storage device, de-identifying read personal data, storing de-identified data in a storage device, and secure deletion of personal data. When the storage device is a NAND flash memory-based storage device, the abovementioned verification operation is performed after secure deletion. Here, secure deletion can be applied in various ways, such as a scrubbing technique, a partial overwrite technique, and a down-bit program technique. When secure deletion does not pass as a verification result, secure deletion may be performed by applying the same deletion technique. Meanwhile, when secure deletion does not pass as a verification result, secure deletion may be performed by applying another deletion technique. The proposed verification technique is intended to be used for the de-identification operation. However, this verification technique is limited to this operation alone. A limited verification technique can be used for any typical deletion operation. That is, it can be used to verify whether a file deleted by a user exists in NAND flash memory. This secure deletion verification technique can be developed into an anti-forensic application. As described above, NAND flash memory receiving an anti-forensic request or an anti-forensic command may XOR scan personal data within all physical blocks and output the result. In other words, it can be applied to check for the existence of personal data.

De-identification of personal information for use in big data operations is an essential requirement. In a NAND flash memory-based storage device, it must be verified that personal information has been completely deleted during the de-identification process. Our research is the first answer this need, and though it addresses the bulk of the issue, there are still areas that need improvement.

*F. Page Buffer without XOR Logic and Verification using Cell Count Information*

The verification technique for secure deletion described above use personal data as a comparison target. However, strictly speaking, it is possible to verify the result of secure deletion without directly comparing personal data. In general, in NAND flash memory, the number of cell counts per page is equally distributed during a normal program operation. For example, in a 3-bit program operation, there are eight threshold voltage distributions, and the number of eight threshold voltage distributions per page is uniformly programmed [42, 43]. Considering this assumption, since secure deletion is a different form from this normal program operation, NAND flash memory outputs a cell count that does not exhibit uniform distribution characteristics. If this assumption is correct, the verification operation for secure deletion can be easily completed using cell count information without a direct comparison of personal data. That is, complete deletion that destroys personal information corresponds to outputting non-uniform cell count information.

On the other hand, deletion that still contains personal information outputs uniform cell count information. Cell count information, which is the criterion for whether verification passes or fails, can be determined by the chip manufacturer. This scenario is advantageous in that verification of secure deletion can be completed without changing the structure of the existing NAND flash memory.

*G. Future Secure Deletion of IoT devices*

De-identification of personal information basically includes secure deletion. For example, de-identification operation for IoT devices, without exception, needs to be equipped with a secure erasure algorithm by default. As described above, verifying secure erasure requires a lot of time and effort, and many problems must be solved in the verification. Therefore, it is expected that it will be difficult to immediately apply it to IoT products. However, these are issues that must be applied one day. Applying a real-time trim operation to secure deletion in existing products would be the most realistic solution. The real-time trim operation means that, while performing a de-identification operation, secure deletion of original data is performed in real time. This real-time trim operation has been disclosed to be immediately applicable with a few modifications of the existing protocol and a simple process. Therefore, subsequent studies on applicable real-time trim operations need to be continued.

In a recent paper [44], although it was actually deleted from the mobile phone, it is still confirming and announcing problems in forensic IoT devices. The need for security in IoT devices has only just begun [44, 45]. This need, if left unattended, is likely to become a very serious problem in the near future. Our research is thus more urgent, more important, and more useful. Although the real-time trim operation is best for security, it still has a problem of performance degradation. In terms of performance optimization, it is expected that the technology will develop in a way that the real-time trim operation is selectively or optionally applied rather than essential. What is clear is that NAND flash memory inevitably requires management of the invalidation area. In-depth research and efforts on such management are needed more and more in the field. This paper presents a method for verifying secure deletion according to de-identification. This is only an example, and research on verification in various ways should be conducted in the near future. In particular, with respect to the demand-based caching technique that maximizes the performance of the SSD, it will be a good research topic to investigate the correlation with the real-time trim technique [46, 47].

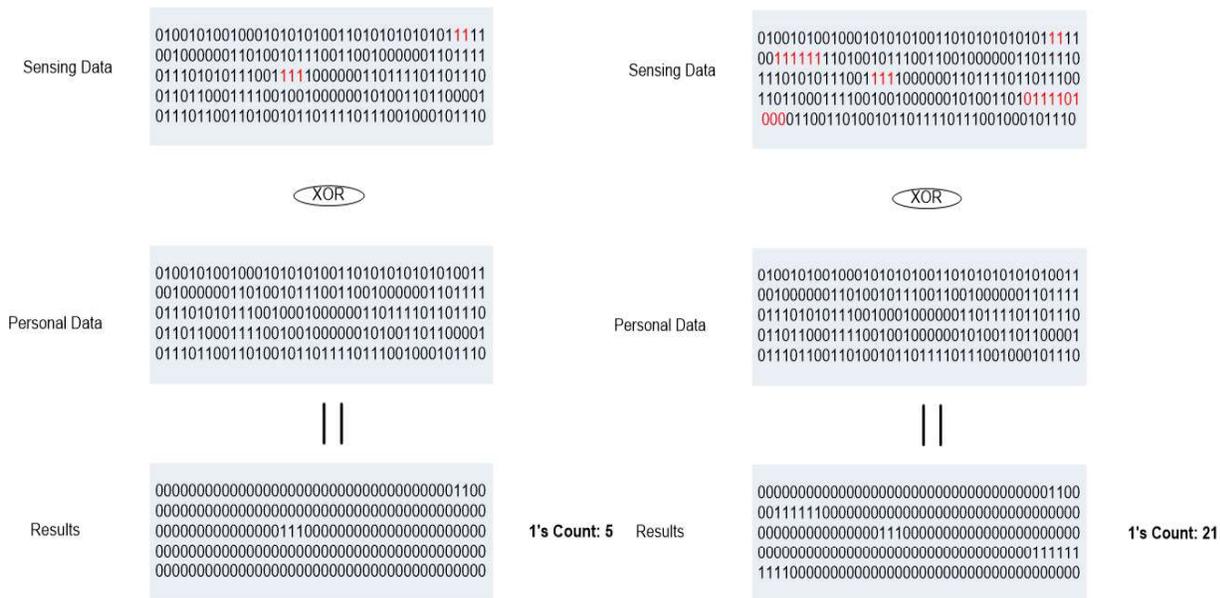

**FIGURE 11.** Verification results using cell count value for secure deletion. (a) In the case of verification failure (Number of 1s: 5), (b) In the case of verification success (Number of 1s: 21).

## VI. VERIFICATION RESULTS & PERFORMANCE EVALUATION

Below, we describe the result of the verification operation according to the XOR comparison. For convenience of explanation, sensing data for 64 bits and personal data were compared. It is assumed that 64 bits are a sector unit. Fig. 11 (a) conceptually shows the verification failure of the secure deletion technique. As a result of the XOR operation of the sensed data and personal data, the counting value of bit "1" is five. When the sensing data and personal data are compared, it means that five bits are different. In this case, the error correction circuit inside NAND flash memory or the error correction circuit outside NAND flash memory is likely to be error corrected. Therefore, the secure deletion technique should indicate verification failure. After this, the deletion operation for the corresponding page should be performed by applying the secure deletion technique again.

Fig. 11 (b) shows that the secure deletion technique has been completely performed. As a result of XORing the sensing data and personal data, the number of bit "1" becomes 21. In this case, the possibility of correction by the error correction circuit is slim. Therefore, the secure deletion operation is considered a verification pass. Success/failure of the verification operation is determined according to the cell count value of 1 for one sector. If the number of '1' count is more than the reference value, secure deletion has been performed successfully. In other words, it means that there is no personal data remaining in the memory block. On the other hand, if the number of counters of 1 is less than the reference value, secure deletion has not performed properly, meaning that personal data still exists inside the memory block. In this case, personal data must be deleted using either a new secure deletion technique or using the same technique again.

Our proposed verification technique operates in real-time. After performing a secure deletion operation, our proposed verification operation can be performed immediately. In general, data to which a randomization technique is applied in NAND flash memory is stored by a program operation. The verification of data deleted by the secure deletion operation can be easily performed only with the distribution characteristic of the cell count. Accordingly, unless there is an even distribution of cell counts, secure deletion is complete. Accordingly, a technique that does not use a XOR operation enables secure deletion verification with only simple cell count information. This verification can be performed only by a read operation. If the count information according to the result of this read operation is much larger than the reference value, secure deletion is deemed successful. On the other hand, if the count information according to the result of this read operation exists in the vicinity of the reference value, secure deletion is considered to have failed.

As mentioned earlier, the part in which we were interested and developed our logic was the original data stored in the unmanaged block due to the de-identification operation. However, in fact, forensics about data stored in the over-provisioning OP area can be a good research theme. All types of memories, including NAND flash memory, contain over-provisioning areas for the purpose of improving performance. For faster processing, it is likely that the original data will be left in the over-provisioning area. Forensic research in this area is part of our future research interests.

TABLE I
PERFORMANCE DIFFERENCES OF SECURE DELETION SCHEMES

| Secure Deletion Scheme | Data Generation | Endurance | Program Disturbance | Verification Reporting |
|---|---|---|---|---|
| Scrubbing [34, 35] | Zero Bit | Increase in cell wear | High | None |
| Partial Overwriting [15, 17] | Possible Random Bit | None | Medium | Possible (Valid Area) |
| Down-Bit Programming [36, 17] | SLC data bit | None | Low | Possible (Valid Area) |
| Deletion Pulse Application [17] | None | None | Low | None |
| Proposed Secure Deletion | Optionally | Optionally | Optionally | P/F Reporting (Valid/Invalid) |

Below, we compared the performance of secure deletion schemes referring to table I. Basically, secure schemes are based on overwriting, so random data must be generated. The scrubbing scheme generates the most significant bit, that is, the zero bit. In the partial overwriting scheme, it is necessary to generate arbitrary data in a programmable state in comparison with the state of the programmed data according to the multi-level cell. The down-bit programming scheme must generate data by converting multi-level cell data into single-level cells. In this case, the down-bit program method requires separate management so that the storage capacity does not change. The deletion pulse application scheme applies a plurality of deletion pulses rather than programming standardized data. This scheme can eliminate the time required for data creation compared to other on-chip secure deletion schemes.

Next, the index related to durability is compared as follows. Since the scrubbing scheme is programmed in the highest state, the wear rate of the cell is inevitably increased significantly. It may be better in terms of durability to perform an erase operation in block units rather than frequently using the scrubbing scheme. The partial overwriting scheme, the down-bit programming scheme, and the deletion pulse application scheme do not have durability issues. Since these secure deletion schemes are basically performed based on overwriting, they inevitably cause program disturbance. Since the scrubbing scheme writes data in the highest state, it has no choice but to proceed to a higher program level. Accordingly, the scrubbing scheme may cause a relatively high program disturbance. It is highly likely that data management of adjacent valid pages will be additionally performed. Similar to the scrubbing scheme, the partial overwriting scheme has the potential to proceed with a program at a relatively high level, and thus inevitably causes program disturbance. However, the partial overwriting scheme causes moderate program disturbance because it is not the data in the highest state of the scrubbing scheme. The down-bit programming scheme and the deletion pulse application scheme cause relatively low program disturbance compared to the scrubbing scheme and the partial overwriting scheme.

Finally, the proposed secure deletion scheme may optionally use at least one of the above mentioned scrubbing, partial-overwriting, down-bit programming, and deletion pulse application. The proposed secure deletion adds a verification scheme to the existing secure deletion schemes. Accordingly, data generation, endurance, and program disturbance of the proposed scheme optionally depend on the selected secure deletion scheme.

Basically, overwriting and program operation include loading target data into page buffers and verifying whether the program operation is properly performed. Therefore, partial overwriting and down-bit programming can be reported according to the verification operation result. However, this verification operation is performed only for a valid area. Existing scrubbing and deletion pulse application are silent about performing the verification operation. That is, they cannot repot result verification result of secure deletion. On the other hand, the proposed secure deletion scheme may compare the data read from the physical area with the original data, and report the verification result according to the comparison result. Such deletion technique is applicable both the valid area and the invalid area.

Also, the proposed secure deletion scheme can verify existing secure deletions, for example, TRIM. In the future, this technical feature can enable real-time TRIM to storage devices in the data servers. In addition, the proposed technique can be used to arbitrarily verify the existence of original data in the storage device according to a user's request. Recently, Intel disclosed DMA algorithm for the data sanctuary according to data movement [45]. The proposed technique will be available for evaluation of such data sanctification algorithms.

## VII. CONCLUSION

When we perform the de-identification operation in IoT devices based on NAND flash memory devices, we confirm the existence of personal data that inevitably remains in the unmanaged block. A complete de-identification operation requires secure deletion of this personal data using various techniques. NAND flash memory has an error correction function to improve data reliability, and this function specifically requires a verification operation for secure deletion. Our proposed verification technique for secure deletion compares personal data and data sensed from unmanaged blocks in 1:1 matching, and discriminates using cell count information from the comparison results. This method makes it possible to perform the verification operation completely in real time without significantly changing the structure of NAND flash memory. As the IoT environment expands further, the forensic issues of NAND flash memory will grow concurrently depending on the existence of original data. This study is expected to be a small but very useful aid to those fighting against privacy-invading forces. Further detailed analyses and experimental studies of this research topic will be necessary in the near future.

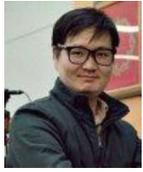

**Na Young Ahn** is a post-doc researcher with the Institute of Cyber Security & Privacy at Korea University, South Korea. He holds a Ph.D. in Cyber Security. He received his B.S. and M.S. degrees from the Department of Electrical Engineering at Korea University. He has been a patent engineer at patent and law firms since 2005. His articles have been published in journals including IEEE Access and Ad Hoc & Sensor Wireless Networks. His research interests include physical layer security in vehicular communications, biometric authentications, Non-Competitive blockchain, and anti-forensics in flash memories.

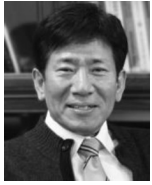

**Dong Hoon Lee** received his B.S. degree in economics from Korea University, Seoul, Korea, in 1985 and M.S. and Ph.D. degrees in computer science from the University of Oklahoma, Norman, OK, USA, in 1988 and 1992, respectively. Since 1993, he has been with the Faculty of Computer Science and Information Security, Korea University. His research interests include the design and analysis of cryptographic protocols in key agreement, encryption, signatures, embedded device security, and privacy-enhancing technology.